\documentclass[a4paper,aps,prd,10pt,preprintnumbers,showpacs,twocolumn,superscriptaddress,nofootinbib]{revtex4-1}
\usepackage{amsmath,amssymb,gensymb,amsfonts}
\usepackage{graphicx}
\usepackage{cmap}
\usepackage[utf8]{inputenc}
\usepackage[T1]{fontenc}
\usepackage{nicefrac}

\begin{document}
\title{Long-Lived Quasinormal Modes and Quasi-Resonances around Non-Minimal Einstein–Yang–Mills Black Holes}
\author{Alexey Dubinsky}
\email{dubinsky@ukr.net}
\affiliation{University of Seville, Seville, Spain}

\begin{abstract}
Using accurate computational methods, we compute the quasinormal frequencies of a massive scalar field propagating near a black hole in the framework of non-minimal Einstein–Yang–Mills theory with a non-zero cosmological constant. We show that increasing the mass of the scalar field significantly decreases the damping rate of the quasinormal modes for both asymptotically flat and de Sitter black holes. However, in the de Sitter case, arbitrarily long-lived modes can exist, whereas in the asymptotically flat case, the damping rate never vanishes completely. In the limit of quasi-resonances, we observe a kind of universal behavior where the frequencies do not depend on the coupling constant. Applying the time-domain integration of perturbation equations we show that even when the effective potential has a negative gap, the scalar field is stable and the perturbations decay in time. In the regime of large mass of the field we obtain the analytic formula for quasinormal modes.
\end{abstract}
\maketitle

\section{Introduction}

The advent of gravitational wave astronomy, inaugurated by the LIGO and Virgo collaborations~\cite{ligo2,ligo1,ligo4,ligo3,ligo5}, has enabled unprecedented access to the strong-field dynamics of black holes. Central to this regime is the so-called ringdown phase, the stage following a merger, during which the system radiates gravitational waves governed by its characteristic quasinormal modes (QNMs)~\cite{review2,review4,review1,review3}. These modes serve as spectral fingerprints of the black hole geometry and offer a model-independent means of probing the near-horizon structure of compact objects.

In parallel, substantial progress has been made in the electromagnetic domain: the Event Horizon Telescope (EHT) has achieved horizon-scale imaging of supermassive black holes~\cite{Goddi:2016qax,EventHorizonTelescope:2019ggy,EventHorizonTelescope:2019dse}, while multiwavelength campaigns have uncovered detailed properties of accretion flows and jets~\cite{Bambi:2015kza}. The combined potential of gravitational and electromagnetic data—an approach known as multi-messenger black hole spectroscopy—has opened a path toward stringent tests of general relativity and its potential ultraviolet completions.

Traditionally, it was thought that QNMs are predominantly shaped by the effective potential's peak, located at intermediate distances from the event horizon. However, recent investigations~\cite{Giesler:2024hcr,Giesler:2019uxc,Konoplya:2022pbc} have demonstrated that early ringdown (or in the frequency domain the overtone sector) can exhibit extreme sensitivity to localized modifications of the metric in the immediate vicinity of the horizon. This sensitivity manifests in a rapid shift of frequencies with seemingly minor geometric changes. 

This realization has spurred extensive investigation into the quasinormal spectra of modified gravity models~\cite{Zinhailo:2024jzt,Konoplya:2024lch,Stuchlik:2025ezz,Bolokhov:2023bwm,Konoplya:2023ahd,Zinhailo:2024kbq,Zhang:2024nny,Stashko:2024wuq}, including those with higher-curvature corrections, non-minimal matter couplings, or higher-dimensional effects. Among these frameworks, non-minimal Einstein–Yang–Mills (EYM) gravity provides a compelling scenario. Its black hole solutions~\cite{Balakin:2015gpq,Balakin:2006gv} resemble Schwarzschild geometry at large radii but feature significant deviations near the horizon, where the QNM sensitivity is most pronounced.

Previous studies~\cite{Lutfuoglu:2025ljm,Gogoi:2024vcx} have examined scalar perturbations in this background for massless fields. However, the role of scalar field mass in this setting remains largely unexplored. Introducing a non-zero field mass not only alters the effective potential—changing its asymptotic behavior and barrier structure—but also leads to novel dynamical phenomena such as the emergence of quasiresonances~\cite{Konoplya:2004wg,Ohashi:2004wr,Konoplya:2017tvu}. These long-lived modes, appearing at specific mass thresholds, are of particular interest for both theoretical and observational reasons, as they may mimic bound states and leave distinctive imprints in the late-time gravitational waveform.

Massive QNMs have generated wide interest across diverse contexts. In brane-world models, mass terms arise naturally from dimensional reduction~\cite{Seahra:2004fg}, while in theories of massive gravity, massive gravitons may contribute to low-frequency signals probed by experiments like the Pulsar Timing Array~\cite{Konoplya:2023fmh,NANOGrav:2023hvm}. The appearance of quasiresonances has been confirmed in numerous setups, including scalar, vector, and tensor fields~\cite{Konoplya:2017tvu,Fernandes:2021qvr}, alternative geometries~\cite{Zinhailo:2018ska}, and even non-traditional objects like wormholes~\cite{Churilova:2019qph}. Moreover, massive fields exhibit oscillatory late-time tails rather than the standard power-law decay, marking another key observational distinction~\cite{Koyama:2001qw,Jing:2004zb,Moderski:2001tk,Konoplya:2006gq,Gibbons:2008rs,Dubinsky:2024jqi}. Mass-like terms can also be induced for otherwise massless fields in the presence of external fields or media~\cite{Kokkotas:2010zd,Konoplya:2007yy,Wu:2015fwa,Davlataliev:2024mjl}. Nonetheless, quasiresonances do not arise universally; for certain spacetimes and parameter regimes, these long-lived states are absent~\cite{Zinhailo:2024jzt}, underscoring the importance of exploring their presence in specific gravitational settings.

In this work, we perform a detailed numerical study of the QNMs of a massive scalar field in the non-minimal EYM black hole background, using the Leaver method~\cite{Leaver:1985ax,Leaver:1986gd} to obtain highly accurate results. We analyze the dependence of the frequencies on the scalar field mass and on the Yang–Mills coupling parameter $\xi$, with a particular focus on overtone behavior at low multipoles. Our results show that the scalar mass suppresses the imaginary part of the frequency, producing long-lived modes, while enhancing the deviations caused by the non-minimal coupling. These effects are especially pronounced in the low-frequency regime and further reinforce the importance of overtones in probing near-horizon physics.

The paper is organized as follows. In Sec.~\ref{sec:background}, we briefly review the black hole solutions in non-minimal Einstein–Yang–Mills theory. In Sec.~\ref{sec:methods}, we describe the numerical methods used for computing the quasinormal modes in both asymptotically flat and asymptotically de Sitter cases. The main results are presented in Sec.~\ref{sec:results}, where we analyze the behavior of the quasinormal frequencies for various values of the scalar field mass, multipole number, and coupling parameters. We also discuss the appearance of quasiresonances and the structure of the effective potential. Finally, in Sec.~\ref{sec:conclusion}, we summarize our findings and outline possible directions for future research.

\section{Black Hole Metric and wave equation}\label{sec:background}

\begin{figure*}
\resizebox{\linewidth}{!}{\includegraphics{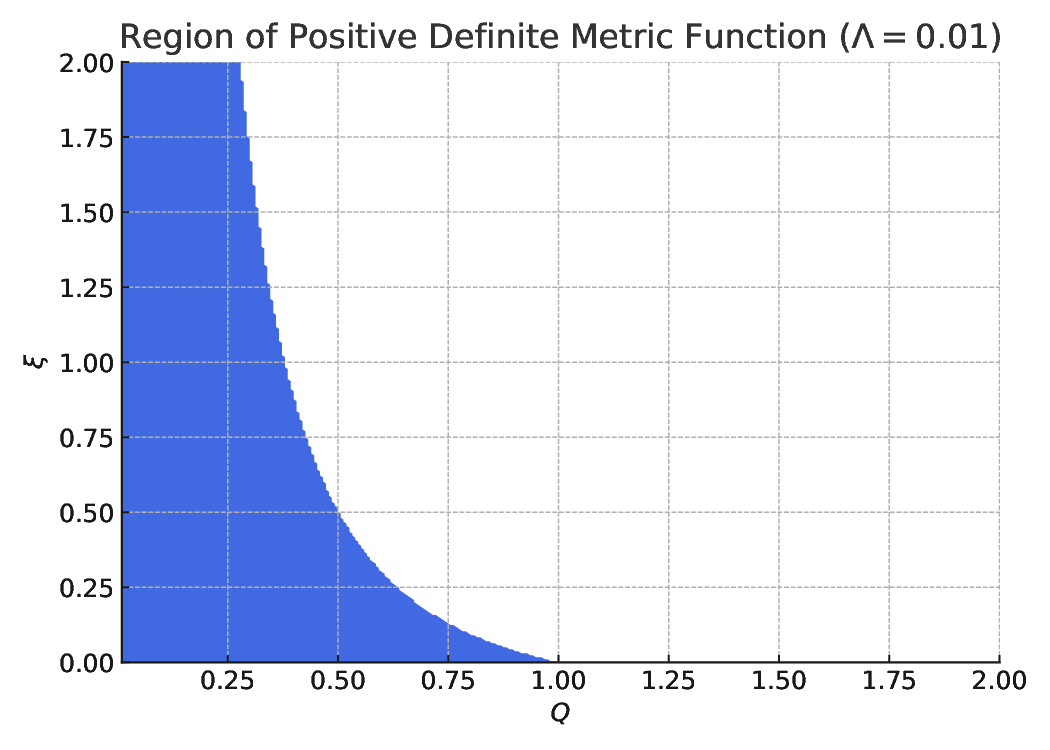}\includegraphics{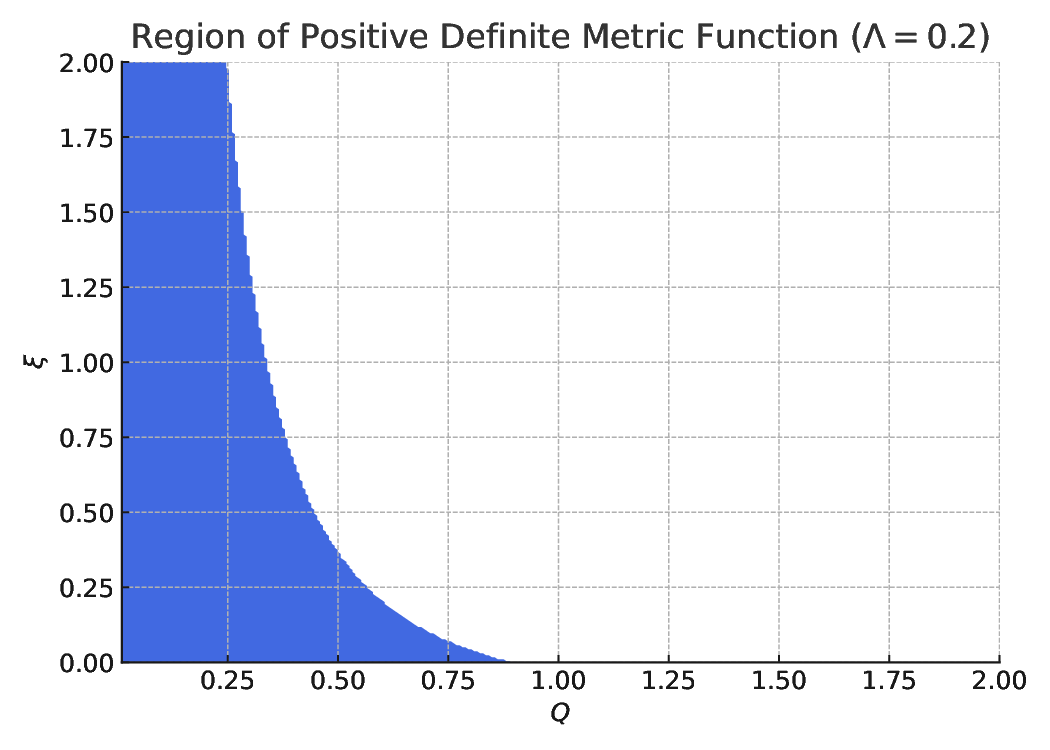}}
\resizebox{\linewidth}{!}{\includegraphics{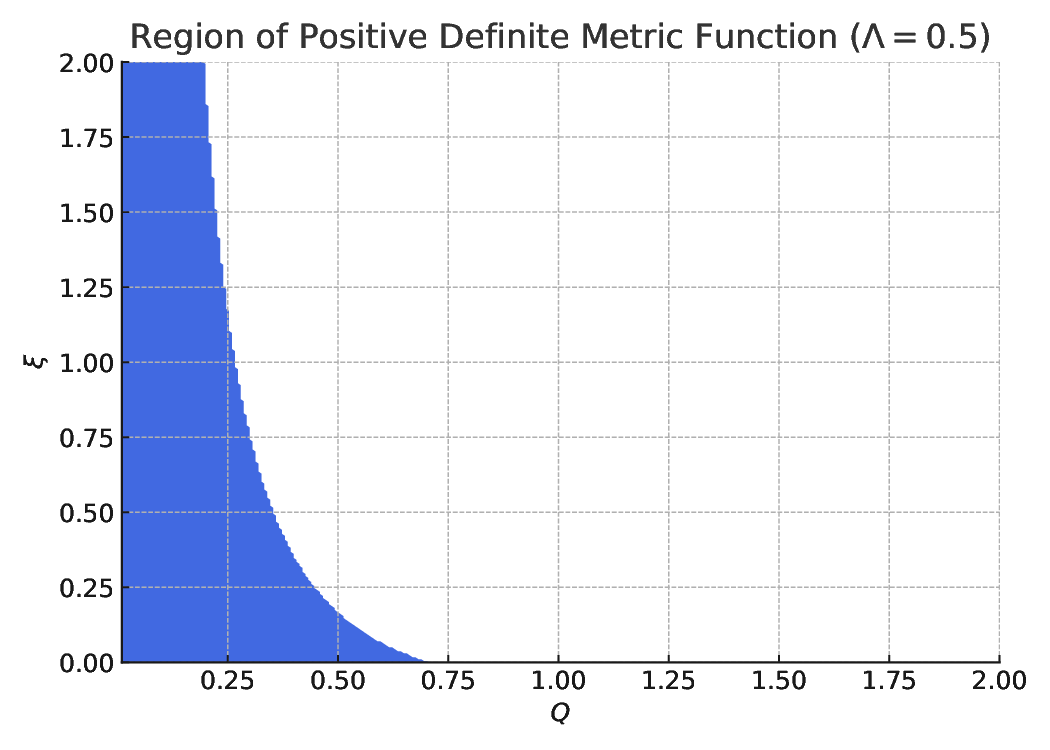}\includegraphics{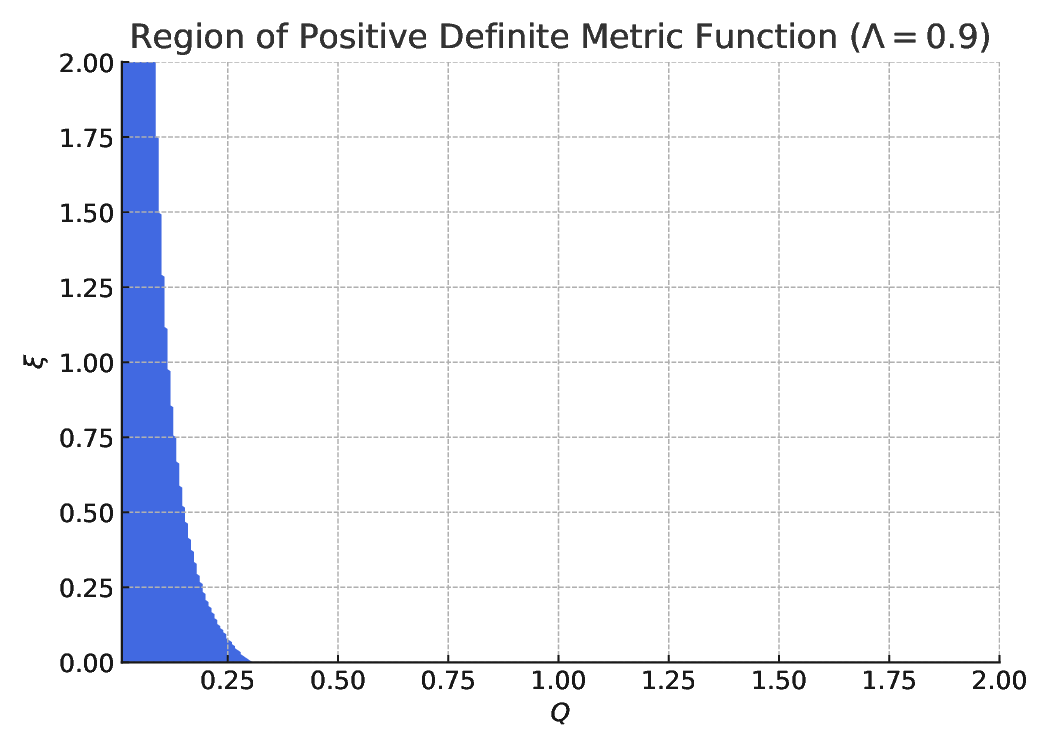}}
\caption{The parametric range allowing for existence of a black hole for various values of $Q$ and $\xi$; $r_{h}=1$.}\label{fig:BHrange}
\end{figure*}

\begin{figure*}
\resizebox{\linewidth}{!}{\includegraphics{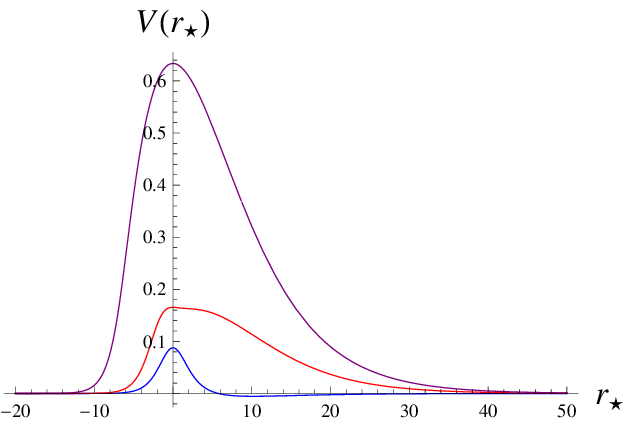}~~\includegraphics{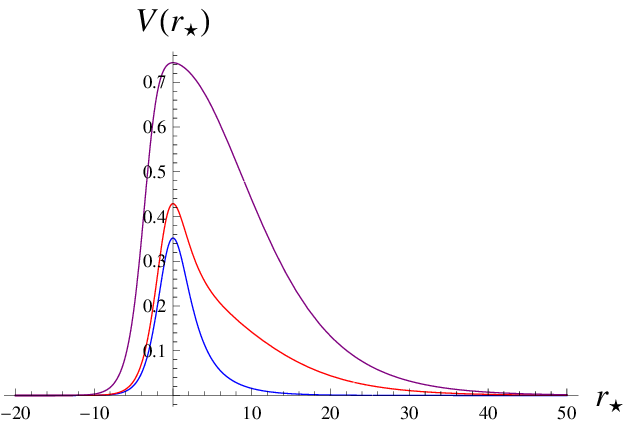}}
\caption{Effective potentials for $\ell=0$ (left) and $\ell=1$ (right) perturbations at $Q=0.2$ $\xi=0.3$ $\Lambda=0.02$ $\mu=0$ (blue line) $\mu=0.5$ (red line), $\mu=1$ (purple line).}\label{fig:Potentials}
\end{figure*}

Interactions between gauge fields and the gravitational sector that extend beyond minimal coupling frequently emerge in a broad range of theoretical contexts. These include semiclassical gravity, where such terms represent quantum backreaction; effective field theories, which accommodate higher-order curvature corrections; and low-energy limits of string theory, where non-minimal structures are common. A prominent example is the Einstein–Yang–Mills (EYM) theory with non-minimal coupling, where the Yang–Mills field interacts explicitly with the spacetime curvature, often through direct coupling to the Ricci scalar. These couplings enrich the solution space of black holes, introducing deviations from classical metrics and providing fertile ground for probing gravitational phenomena beyond general relativity.

In this work, we study black hole solutions arising from a generalized non-minimal EYM theory with a cosmological constant. The action is given by:
\begin{equation}
\mathcal{S} = \int d^4x \sqrt{-g} \left[ \frac{R - 2\Lambda}{16\pi G} - \frac{1}{4} F^a_{\mu\nu} F^{a\mu\nu} + \frac{\xi}{4} R F^a_{\mu\nu} F^{a\mu\nu} \right].
\end{equation}
Here, \( R \) denotes the Ricci scalar, \( \Lambda \) is the cosmological constant, and \( F^a_{\mu\nu} \) represents the field strength tensor of the Yang–Mills field associated with an \( SU(2) \) gauge group. The dimensionless parameter \( \xi \) characterizes the strength of the non-minimal coupling between the gauge sector and spacetime curvature. The non-minimal interaction term, proportional to \( \xi R F^2 \), introduces curvature-dependent dynamics, making the behavior of the resulting solutions particularly sensitive to both the sign and magnitude of \( \Lambda \).

This theoretical framework admits static, spherically symmetric black hole solutions, where the Yang–Mills field is realized through a Wu–Yang-type purely magnetic monopole configuration. The corresponding spacetime metric is given by
\begin{equation}
ds^2 = -f(r)\,dt^2 + \frac{dr^2}{f(r)} + r^2(d\theta^2 + \sin^2\theta\,d\phi^2),
\end{equation}
with the lapse function \( f(r) \) taking the form
\begin{equation}
f(r) = 1 + \left( \frac{r^4}{r^4 + 2\xi Q^2} \right) \left( \frac{Q^2}{r^2} - \frac{2M}{r} - \frac{\Lambda r^2}{3} \right),
\end{equation}
where \( M \) is the ADM mass of the black hole, \( Q \) denotes the Yang–Mills magnetic charge, and \( \xi \) governs the degree of non-minimal interaction between geometry and gauge fields. In the limit \( \xi = 0 \), the metric reduces to that of the minimally coupled Einstein–Yang–Mills system with a cosmological constant. Setting \( Q = 0 \) further recovers the Schwarzschild–(anti)–de Sitter geometry.

The presence of \( \Lambda \) modifies the asymptotic structure of the spacetime. For \( \Lambda = 0 \), the solution is asymptotically flat and suitable for standard quasinormal mode boundary conditions. For \( \Lambda > 0 \), the black hole is embedded in an expanding de Sitter universe, and the spacetime contains a cosmological horizon in addition to the event horizon. This changes the behavior of perturbations and the allowed quasinormal spectra.

To investigate the dynamical response of these black holes, we consider a massive scalar test field governed by the covariant Klein–Gordon equation:
\begin{equation}
\frac{1}{\sqrt{-g}} \partial_\mu \left( \sqrt{-g}\,g^{\mu\nu} \partial_\nu \Phi \right) = \mu^2 \Phi,
\end{equation}
where \( \mu \) is the mass of the scalar field. Assuming a separable ansatz,
\[
\Phi(t, r, \theta, \phi) = e^{-i \omega t} \frac{\Psi(r)}{r} Y_{\ell m}(\theta, \phi),
\]
one obtains a radial equation in Schrödinger-like form:
\begin{equation}
\frac{d^2\Psi}{dr_*^2} + \left( \omega^2 - V(r) \right)\Psi = 0,
\end{equation}
where the tortoise coordinate is defined by \( dr_*/dr = 1/f(r) \), and the effective potential reads:
\begin{equation}
V(r) = f(r) \left[ \mu^2 + \frac{\ell(\ell+1)}{r^2} + \frac{f'(r)}{r} \right].
\end{equation}
The inclusion of the mass term \( \mu \) and the cosmological constant \( \Lambda \) modifies both the shape and the asymptotic behavior of the potential, which can significantly affect the quasinormal spectrum. In the asymptotically flat case, the potential decays to zero at spatial infinity, while in the de Sitter case it vanishes at the cosmological horizon, requiring adapted boundary conditions.

These modifications lead to new dynamical features, including quasiresonant behavior for large \( \mu \), suppressed damping for low overtones, and possible sensitivity of the spectrum to near-horizon geometry. The goal of this work is to explore these effects across both asymptotically flat and de Sitter branches of the non-minimal EYM black hole family.

The metric function describes a black hole when it is positive definite between the event horizon \( r_h \) and the cosmological horizon \( r_c \), and vanishes at both boundaries. The parameter range that allows for a black hole solution is shown in Fig.~\ref{fig:BHrange}. In fig. \ref{fig:Potentials} one can see that the effective potential is positive from the event to the cosmological horizon (situated at infinity in terms of the tortoise coordinate), except for the case $\ell=0$ which allows for a negative gap. Thus, the stability of the $\ell=0$ perturbations is not guaranteed and must be checked via the analysis of quasinormal modes. 

\begin{figure*}
\resizebox{\linewidth}{!}{\includegraphics{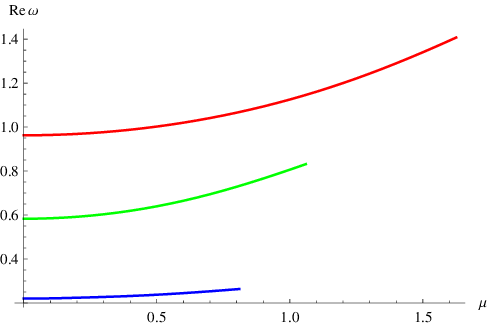}~~\includegraphics{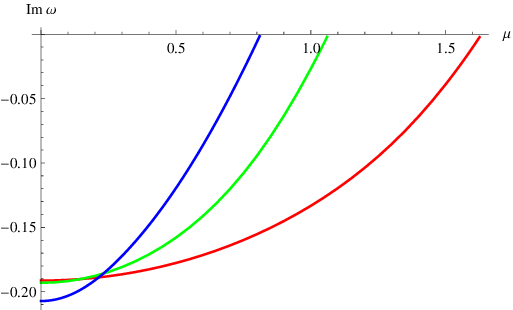}}
\caption{Real and imaginary parts of QNMs for $n=0$, $\ell=0$ (blue), $1$ (green), $2$ (red); $\xi=0.1$, $Q=0.1$, $r_{h}=1$.}\label{fig:QNM1}
\end{figure*}

\begin{figure*}
\resizebox{\linewidth}{!}{\includegraphics{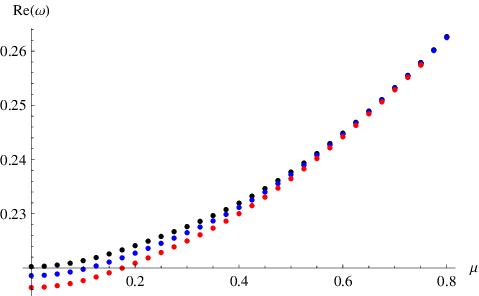}~~\includegraphics{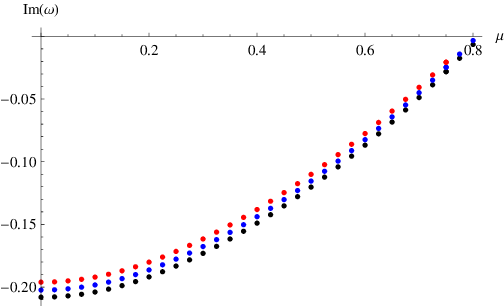}}
\caption{Real and imaginary parts of QNMs for $n=0$, $\ell=0$, $Q=0.1$, $r_{h}=1$: $\xi=0$ (black), $\xi=0.7$ (blue), $\xi=1.5$ (red).}\label{fig:QNM2}
\end{figure*}

\section{Methods for calculations of QNMs}\label{sec:methods}

\subsection{Time-Domain Integration and Frequency Extraction}
\label{sec:timedomain}

For a massive field, this asymptotic condition is modified compared to the massless case, since the field does not propagate to infinity as a free wave but decays as an evanescent mode:
\begin{equation}\label{bcond}
\Psi(r) \sim 
\begin{cases}
e^{-i \omega r_*}, & r \to r_h \quad (\text{horizon}), \\
e^{\sqrt{\omega^2 - \mu^2 }\, r_{*}}, & r \to \infty,
\end{cases}
\end{equation}
where \( \mu \) is the mass of the scalar field, and \( r_* \) is the tortoise coordinate defined by \( dr_*/dr = 1/f(r) \).

An alternative to frequency-domain techniques such as the Frobenius or WKB methods is the time-domain integration approach, which provides a direct numerical solution of the wave equation in the time coordinate. This method, first developed by Gundlach, Price, and Pullin~\cite{Gundlach:1993tp}, is particularly valuable when the effective potential is known only numerically, or when one seeks to study the nonlinear evolution of perturbations or the excitation of multiple modes simultaneously.

The basic idea is to numerically evolve the scalar field perturbation governed by the wave equation:
\begin{equation}
\frac{\partial^2 \Psi(t, r_*)}{\partial t^2} - \frac{\partial^2 \Psi(t, r_*)}{\partial r_*^2} + V(r)\Psi(t, r_*) = 0,
\end{equation}
where \( r_* \) is the tortoise coordinate defined by \( dr_*/dr = 1/f(r) \), and \( V(r) \) is the effective potential. This hyperbolic partial differential equation is typically integrated using characteristic coordinates:
\[
u = t - r_*, \qquad v = t + r_*,
\]
which naturally align with ingoing and outgoing null rays. In these coordinates, the wave equation simplifies to:
\begin{equation}
4 \frac{\partial^2 \Psi}{\partial u \partial v} + V(r(u,v)) \Psi = 0.
\end{equation}

The numerical domain is discretized on a grid of points in the \( (u,v) \) plane, and the evolution proceeds using a finite difference scheme. A widely used stable second-order scheme is:
\begin{equation}
\Psi_N = \Psi_W + \Psi_E - \Psi_S - \Delta^2 \frac{V_S}{8} (\Psi_W + \Psi_E),
\end{equation}
where the points \( N, W, E, S \) form a diamond-shaped stencil on the grid. The potential \( V_S \) is evaluated at the south point. The initial data are typically specified on the two null surfaces \( u = u_0 \) and \( v = v_0 \), for instance using a Gaussian pulse centered at some \( r_* = r_0 \):
\begin{equation}
\Psi(u_0, v) = \exp\left[ -\frac{(v - v_c)^2}{2\sigma^2} \right], \qquad \Psi(u, v_0) = 0.
\end{equation}

The result of the evolution is the scalar field \( \Psi(t, r_*) \) at a fixed spatial position, typically far from the black hole, as a function of time. After an initial transient phase, the solution exhibits an exponentially damped oscillatory behavior characteristic of quasinormal ringing:
\begin{equation}
\Psi(t) \sim \sum_{n=0}^\infty A_n e^{-i \omega_n t},
\end{equation}
where \( \omega_n \) are the QNM frequencies and \( A_n \) are the corresponding excitation amplitudes.

To extract the frequencies \( \omega_n \) from the time-domain signal, one can apply the \emph{Prony method}, which fits the numerical data to a sum of damped exponentials:
\begin{equation}
\Psi(t_k) \approx \sum_{n=1}^{N} A_n e^{-i \omega_n t_k}, \quad t_k = t_0 + k\Delta t.
\end{equation}
This method transforms the fitting problem into a generalized eigenvalue problem for frequencies \( \omega_n \), assuming a fixed number of modes \( N \) and uniform time sampling. The Prony method is particularly well-suited to clean, high-resolution time-domain data and can resolve both fundamental modes and higher overtones when properly implemented.

One of the main advantages of the time-domain approach is its ability to capture the full evolution of the perturbation, including the late-time tails and the interplay between different QNMs. Therefore, it was used in a great number of works with proved accuracy and efficiency \cite{Momennia:2022tug,Skvortsova:2023zca,Skvortsova:2024wly,Aneesh:2018hlp,Lutfuoglu:2025hjy,Churilova:2021tgn,Malik:2024tuf,Malik:2024elk}. It is also robust against numerical noise and does not require detailed knowledge of the asymptotic structure of the effective potential. As such, it complements frequency-domain methods and serves as an important tool for cross-validation and for analyzing QNMs in complex or numerically defined spacetimes.

\subsection{The Leaver method}

To compute quasinormal modes of a massive scalar field in asymptotically flat black hole spacetimes, we adopt the Frobenius method developed by Leaver~\cite{Leaver:1985ax}, which has proven to be a highly accurate and reliable semi-analytical technique. In this context, the field satisfies the massive Klein–Gordon equation, and the appropriate QNM boundary conditions demand purely ingoing waves at the event horizon and exponentially decaying solutions at spatial infinity. 

To capture the asymptotic behavior, we express the radial solution as a generalized Frobenius series around the event horizon:
\begin{equation}
\Psi(r) = F(r) \sum_{n=0}^{\infty} a_n \left( \frac{r - r_h}{r} \right)^n,
\end{equation}
where the prefactor $F(r)$ ensures correct boundary conditions (\ref{bcond}) at both the horizon and spatial infinity and regularity of the expansion.

Substituting this ansatz into the radial equation yields a linear recurrence relation for the expansion coefficients \( a_n \). FIn some cases, this recurrence relation has three terms:
\begin{equation}
\alpha_n a_{n+1} + \beta_n a_n + \gamma_n a_{n-1} = 0 \quad (n \geq 1),
\end{equation}
with the initial condition in the form:
\begin{equation}
\alpha_0 a_1 + \beta_0 a_0 = 0.
\end{equation}
The quasinormal frequencies are determined by imposing the convergence condition for the infinite series, which leads to a continued fraction equation:
\[
\frac{a_1}{a_0} = -\frac{\beta_0}{\alpha_0} = 
\frac{\gamma_1}{
\beta_1 - \dfrac{\alpha_1 \gamma_2}{
\beta_2 - \dfrac{\alpha_2 \gamma_3}{
\beta_3 - \dfrac{\alpha_3 \gamma_4}{
\beta_4 - \cdots
}}}}
\]

Solving the above continued fraction numerically yields the allowed QNM frequencies \( \omega \) for the given values of \( \mu \), \( \ell \), and the black hole parameters.

In some cases, the recurrence relation involves more than three terms due to the complexity of the background or the form of the wave equation. Such higher-order relations can be systematically reduced to a three-term recurrence using Gaussian elimination techniques~\cite{Leaver:1986gd}. The resulting simplified form is then suitable for the standard continued fraction analysis.

In practical calculations, the convergence of the continued fraction can deteriorate for high overtones or in the presence of irregular singularities. To address this, we employ two improvements: the method of \textit{integration through the midpoint} ~\cite{Rostworowski:2006bp}, which avoids spurious singularities by evaluating the series at an interior point, and the \textit{Nollert improvemen}t ~\cite{Nollert:1993zz,Zhidenko:2006rs}, which analytically approximates the tail of the continued fraction to accelerate convergence.

This refined Leaver method has been widely applied to quasinormal mode calculations for massless and massive fields across a variety of black hole spacetimes~\cite{Konoplya:2004wg,Ohashi:2004wr,Konoplya:2017tvu,Rosa:2011my,Konoplya:2007zx,Kanti:2006ua,Zinhailo:2024jzt,Dubinsky:2024fvi}, and remains one of the most precise tools for studying both fundamental and overtone modes in black hole perturbation theory.

\subsection{WKB Method with Padé Approximants}\label{sec:wkb}

The Wentzel–Kramers–Brillouin (WKB) approximation provides a powerful semi-analytic technique for estimating quasinormal mode (QNM) frequencies in black hole spacetimes with smooth and localized effective potentials. It is especially well-suited for spacetimes where the potential has a single, positive-definite peak and decays asymptotically—conditions commonly met in asymptotically flat and de Sitter geometries.

For perturbations governed by a Schrödinger-like wave equation,
\begin{equation}
\frac{d^2 \Psi}{dr_*^2} + \left(\omega^2 - V(r)\right) \Psi = 0,
\end{equation}
the WKB method yields an approximate relation between the frequency \( \omega \) and the shape of the potential barrier. The method was suggested by Schutz and Will \cite{Schutz:1985km}, and later extended to sixth and higher orders~\cite{Iyer:1986np,Konoplya:2003ii,Matyjasek:2017psv}. The general \( N \)th-order WKB formula takes the form:
\begin{equation}
i \frac{\omega^2 - V_0}{\sqrt{-2 V_0''}} - \sum_{j=2}^{N} \Lambda_j = n + \frac{1}{2}, \quad n = 0, 1, 2, \dots,
\end{equation}
where \( V_0 \) is the height of the effective potential, \( V_0'' \) is its second derivative with respect to the tortoise coordinate \( r_* \), and \( \Lambda_j \) are correction terms arising from higher-order derivatives of \( V(r) \).

The WKB series is inherently asymptotic and may diverge beyond a certain order, particularly for low multipole numbers \( \ell \) or higher overtones \( n \). To mitigate this issue and improve numerical accuracy, Padé resummation techniques are employed. In this approach, the formal series is replaced by a rational Padé approximant \( \text{Padé}_{\tilde{n}}^{\tilde{m}} \), which often provides superior convergence properties and stability. For sixth-order calculations, one commonly uses \( \tilde{m} = \tilde{n} = 3 \), though other configurations are also possible~\cite{Matyjasek:2017psv}.

The Padé-improved WKB method is particularly effective in backgrounds where the effective potential is smooth and possesses a single maximum \cite{Konoplya:2001ji,Zinhailo:2019rwd,Xiong:2021cth,Skvortsova:2024atk,Hamil:2024nrv,Liu:2024wch,Konoplya:2005sy,Kodama:2009bf,Malik:2025ava,Konoplya:2020jgt,Dubinsky:2024hmn,Lutfuoglu:2025hwh}. For massive scalar fields in asymptotically flat or de Sitter spacetimes, the mass term \( \mu^2 \) enhances the positivity of the effective potential:
\begin{equation}
V(r) = f(r) \left[ \mu^2 + \frac{\ell(\ell+1)}{r^2} + \frac{f'(r)}{r} \right],
\end{equation}
where \( f(r) \) is the black hole metric function. The presence of a cosmological constant \( \Lambda > 0 \) causes the potential to decay to zero near the cosmological horizon, while the scalar mass ensures positivity at large \( r \). Together, these features guarantee a single-peaked, positive-definite potential for a wide range of parameters, making the WKB method well-justified.

Despite its strengths, the WKB approach must be applied with care. In spacetimes where the potential develops additional local extrema or negative gaps — for instance, in higher-curvature gravity, non-minimal couplings, or geometries with quantum corrections — the method may fail to converge or yield unphysical results. In such cases, more robust numerical methods such as Leaver’s continued fraction method or time-domain integration are required to obtain reliable QNM spectra.

Nonetheless, for spacetimes with well-behaved potentials, the WKB method remains a fast and insightful tool, especially useful for cross-validating results and understanding the dependence of QNMs on physical parameters such as mass, charge, and cosmological constant~\cite{Konoplya:2001ji,Zinhailo:2019rwd,Xiong:2021cth,Skvortsova:2024atk}.

\begin{figure*}
\resizebox{\linewidth}{!}{\includegraphics{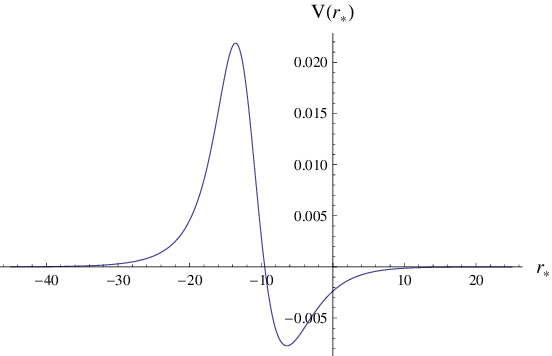}~~\includegraphics{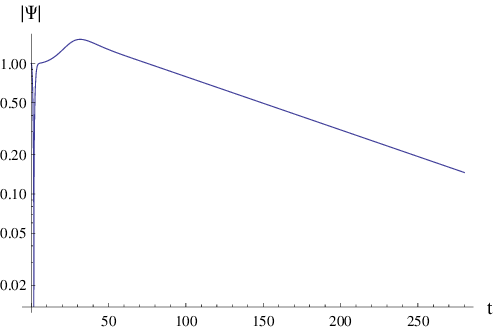}}
\caption{Effective potentials for $\ell=0$  perturbations (left) and time-domain profile for $Q=0.2$, $\xi=2$, $\Lambda=0.2$, $\mu=0.1$, $r_h =1$. The quasinormal mode dominating at asymptotic times belongs to the de Sitter branch: $\omega =  - 0.009397 i$}\label{fig:TD1}
\end{figure*}

\begin{figure*}
\resizebox{\linewidth}{!}{\includegraphics{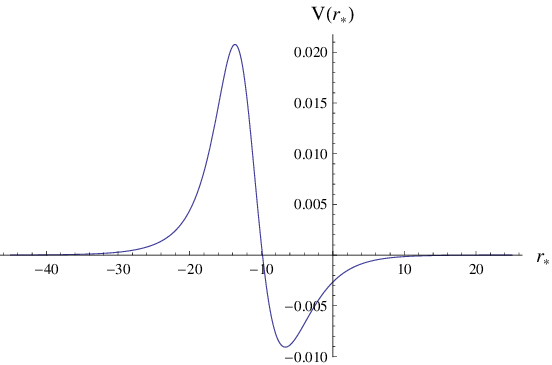}~~\includegraphics{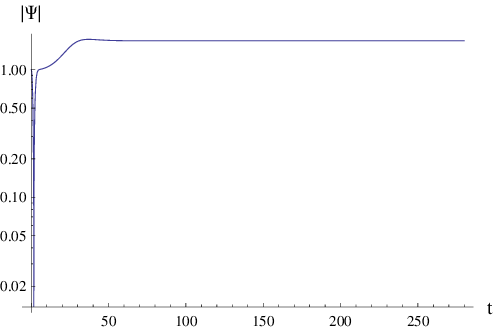}}
\caption{Effective potentials for $\ell=0$  perturbations (left) and time-domain profile for $Q=0.2$, $\xi=2$, $\Lambda=0.2$, $\mu=0$, $r_h =1$. The quasinormal mode dominating at asymptotic times belongs to the de Sitter branch: $\omega = - 1.58411 \cdot 10^{-7} i$}\label{fig:TD2}
\end{figure*}

\begin{tiny}
\begin{table}
\begin{tabular}{c c c c c}
\hline
$\xi$ & $\mu$ & WKB-6 ($m=3$) & WKB-9 ($m=4$) & difference  \\
\hline
$0$ & $0.1$ & $0.221977-0.203817 i$ & $0.220916-0.203705 i$ & $0.354\%$\\
$0$ & $0.5$ & $0.420794-0.035207 i$ & $0.424102-0.038712 i$ & $1.14\%$\\
$0$ & $1.$ & $0.847547-0.041384 i$ & $0.847524-0.041398 i$ & $0.00310\%$\\
$0$ & $1.5$ & $1.270601-0.041903 i$ & $1.270600-0.041901 i$ & $0.00012\%$\\
$0$ & $2.$ & $1.693768-0.042081 i$ & $1.693768-0.042081 i$ & $0.00002\%$\\
$0$ & $2.5$ & $2.116989-0.042163 i$ & $2.116989-0.042163 i$ & $2.3\times 10^{-6}\%$\\
$0$ & $3.$ & $2.540240-0.042207 i$ & $2.540240-0.042207 i$ & $0\%$\\
$0$ & $3.5$ & $2.963509-0.042233 i$ & $2.963509-0.042233 i$ & $0\%$\\
$0.5$ & $0.1$ & $0.219014-0.199905 i$ & $0.220060-0.200445 i$ & $0.397\%$\\
$0.5$ & $0.5$ & $0.421017-0.035814 i$ & $0.423383-0.038755 i$ & $0.893\%$\\
$0.5$ & $1.$ & $0.846423-0.041341 i$ & $0.846401-0.041352 i$ & $0.00284\%$\\
$0.5$ & $1.5$ & $1.268927-0.041851 i$ & $1.268926-0.041850 i$ & $0.00010\%$\\
$0.5$ & $2.$ & $1.691543-0.042027 i$ & $1.691543-0.042027 i$ & $0.00002\%$\\
$0.5$ & $2.5$ & $2.114212-0.042108 i$ & $2.114212-0.042108 i$ & $2.3\times 10^{-6}\%$\\
$0.5$ & $3.$ & $2.536911-0.042151 i$ & $2.536911-0.042151 i$ & $0\%$\\
$0.5$ & $3.5$ & $2.959627-0.042177 i$ & $2.959627-0.042177 i$ & $0\%$\\
$1.$ & $0.1$ & $0.219004-0.197391 i$ & $0.217054-0.196784 i$ & $0.693\%$\\
$1.$ & $0.5$ & $0.421281-0.036498 i$ & $0.422884-0.038726 i$ & $0.649\%$\\
$1.$ & $1.$ & $0.845301-0.041298 i$ & $0.845280-0.041305 i$ & $0.00257\%$\\
$1.$ & $1.5$ & $1.267256-0.041800 i$ & $1.267255-0.041799 i$ & $0.00008\%$\\
$1.$ & $2.$ & $1.689323-0.041974 i$ & $1.689322-0.041974 i$ & $0.00001\%$\\
$1.$ & $2.5$ & $2.111441-0.042053 i$ & $2.111441-0.042053 i$ & $2.3\times 10^{-6}\%$\\
$1.$ & $3.$ & $2.533588-0.042096 i$ & $2.533588-0.042096 i$ & $0\%$\\
$1.$ & $3.5$ & $2.955754-0.042122 i$ & $2.955754-0.042122 i$ & $0\%$\\
$1.5$ & $0.1$ & $0.220577-0.192122 i$ & $0.217081-0.191494 i$ & $1.21\%$\\
$1.5$ & $0.5$ & $0.422034-0.037838 i$ & $0.422213-0.038734 i$ & $0.216\%$\\
$1.5$ & $1.$ & $0.844181-0.041255 i$ & $0.844162-0.041259 i$ & $0.00232\%$\\
$1.5$ & $1.5$ & $1.265589-0.041749 i$ & $1.265588-0.041748 i$ & $0.00006\%$\\
$1.5$ & $2.$ & $1.687107-0.041920 i$ & $1.687106-0.041920 i$ & $0.00001\%$\\
$1.5$ & $2.5$ & $2.108675-0.041999 i$ & $2.108675-0.041999 i$ & $2.3\times 10^{-6}\%$\\
$1.5$ & $3.$ & $2.530273-0.042041 i$ & $2.530273-0.042041 i$ & $0\%$\\
$1.5$ & $3.5$ & $2.951887-0.042066 i$ & $2.951887-0.042066 i$ & $0\%$\\
\hline
\end{tabular}
\caption{Quasinormal modes of the $\ell=0$ test scalar field  for the Yang-Mills black hole ($r_h=1$) $\Lambda=0.001, ~ Q=0.1$  calculated using the WKB formula at different orders and Pade approximants.}\label{table1}
\end{table}
\begin{table*}
\begin{tabular}{c c c c c}
\hline
$\xi$ & $\mu$ & WKB-6 ($m=3$) & WKB-9 ($m=4$) & difference  \\
\hline
$0$ & $0.1$ & $0.580275-0.191664 i$ & $0.580282-0.191660 i$ & $0.00117\%$\\
$0$ & $0.5$ & $0.634411-0.158664 i$ & $0.634463-0.158623 i$ & $0.0103\%$\\
$0$ & $1.$ & $0.897603-0.038230 i$ & $0.897083-0.039413 i$ & $0.144\%$\\
$0$ & $1.5$ & $1.292790-0.039242 i$ & $1.292747-0.039232 i$ & $0.00343\%$\\
$0$ & $2.$ & $1.709611-0.040873 i$ & $1.709613-0.040876 i$ & $0.00019\%$\\
$0$ & $2.5$ & $2.129397-0.041459 i$ & $2.129398-0.041459 i$ & $5.5\times 10^{-6}\%$\\
$0$ & $3.$ & $2.550465-0.041740 i$ & $2.550465-0.041740 i$ & $0\%$\\
$0$ & $3.5$ & $2.972215-0.041899 i$ & $2.972215-0.041899 i$ & $0\%$\\
$0.5$ & $0.1$ & $0.575657-0.188141 i$ & $0.575636-0.188105 i$ & $0.00677\%$\\
$0.5$ & $0.5$ & $0.630223-0.155378 i$ & $0.630154-0.155369 i$ & $0.0108\%$\\
$0.5$ & $1.$ & $0.899253-0.031593 i$ & $0.895583-0.039166 i$ & $0.935\%$\\
$0.5$ & $1.5$ & $1.290909-0.039259 i$ & $1.290867-0.039242 i$ & $0.00353\%$\\
$0.5$ & $2.$ & $1.707249-0.040845 i$ & $1.707251-0.040847 i$ & $0.00017\%$\\
$0.5$ & $2.5$ & $2.126517-0.041418 i$ & $2.126517-0.041418 i$ & $4.4\times 10^{-6}\%$\\
$0.5$ & $3.$ & $2.547052-0.041693 i$ & $2.547052-0.041693 i$ & $0\%$\\
$0.5$ & $3.5$ & $2.968263-0.041850 i$ & $2.968263-0.041850 i$ & $0\%$\\
$1.$ & $0.1$ & $0.571010-0.184714 i$ & $0.570971-0.184662 i$ & $0.0109\%$\\
$1.$ & $0.5$ & $0.626010-0.152171 i$ & $0.625863-0.152179 i$ & $0.0228\%$\\
$1.$ & $1.$ & $0.903983-0.018856 i$ & $0.893933-0.038788 i$ & $2.47\%$\\
$1.$ & $1.5$ & $1.289036-0.039274 i$ & $1.288996-0.039251 i$ & $0.00361\%$\\
$1.$ & $2.$ & $1.704894-0.040816 i$ & $1.704895-0.040818 i$ & $0.00015\%$\\
$1.$ & $2.5$ & $2.123644-0.041376 i$ & $2.123644-0.041376 i$ & $3.6\times 10^{-6}\%$\\
$1.$ & $3.$ & $2.543647-0.041646 i$ & $2.543647-0.041646 i$ & $0\%$\\
$1.$ & $3.5$ & $2.964320-0.041800 i$ & $2.964320-0.041800 i$ & $0\%$\\
$1.5$ & $0.1$ & $0.566338-0.181388 i$ & $0.566289-0.181329 i$ & $0.0129\%$\\
$1.5$ & $0.5$ & $0.621775-0.149053 i$ & $0.621579-0.149087 i$ & $0.0310\%$\\
$1.5$ & $1.$ & $0.894470-0.031586 i$ & $0.892114-0.038408 i$ & $0.806\%$\\
$1.5$ & $1.5$ & $1.287172-0.039286 i$ & $1.287132-0.039259 i$ & $0.00372\%$\\
$1.5$ & $2.$ & $1.702545-0.040786 i$ & $1.702546-0.040788 i$ & $0.00014\%$\\
$1.5$ & $2.5$ & $2.120778-0.041334 i$ & $2.120778-0.041334 i$ & $0\%$\\
$1.5$ & $3.$ & $2.540250-0.041599 i$ & $2.540250-0.041599 i$ & $0\%$\\
$1.5$ & $3.5$ & $2.960385-0.041750 i$ & $2.960385-0.041750 i$ & $0\%$\\
\hline
\end{tabular}
\caption{Quasinormal modes of the $\ell=1$ test scalar field  for the Yang-Mills black hole ($r_h=1$) $\Lambda=0.001, ~ Q=0.1$ calculated using the WKB formula at different orders and Pade approximants.}\label{table2}
\end{table*}
\begin{table}
\begin{tabular}{c c c c c}
\hline
$Q$ & $\mu$ & WKB-6 ($m=3$) & WKB-9 ($m=4$) & difference  \\
\hline
$0$ & $0.1$ & $0.222726-0.205603 i$ & $0.221653-0.205472 i$ & $0.357\%$\\
$0$ & $0.5$ & $0.420693-0.034807 i$ & $0.424634-0.038705 i$ & $1.31\%$\\
$0$ & $1.$ & $0.848467-0.041445 i$ & $0.848443-0.041461 i$ & $0.00339\%$\\
$0$ & $1.5$ & $1.271967-0.041970 i$ & $1.271966-0.041969 i$ & $0.00013\%$\\
$0$ & $2.$ & $1.695581-0.042150 i$ & $1.695581-0.042150 i$ & $0.00002\%$\\
$0$ & $2.5$ & $2.119251-0.042233 i$ & $2.119251-0.042233 i$ & $2.3\times 10^{-6}\%$\\
$0$ & $3.$ & $2.542952-0.042278 i$ & $2.542952-0.042278 i$ & $0\%$\\
$0$ & $3.5$ & $2.966671-0.042304 i$ & $2.966671-0.042304 i$ & $0\%$\\
$0.2$ & $0.1$ & $0.217151-0.195462 i$ & $0.218031-0.195966 i$ & $0.347\%$\\
$0.2$ & $0.5$ & $0.421176-0.036987 i$ & $0.422089-0.038666 i$ & $0.452\%$\\
$0.2$ & $1.$ & $0.843894-0.041168 i$ & $0.843877-0.041170 i$ & $0.00199\%$\\
$0.2$ & $1.5$ & $1.265174-0.041661 i$ & $1.265174-0.041661 i$ & $0.00005\%$\\
$0.2$ & $2.$ & $1.686559-0.041833 i$ & $1.686559-0.041833 i$ & $9.0\times 10^{-6}\%$\\
$0.2$ & $2.5$ & $2.107994-0.041912 i$ & $2.107994-0.041912 i$ & $2.9\times 10^{-6}\%$\\
$0.2$ & $3.$ & $2.529456-0.041955 i$ & $2.529456-0.041954 i$ & $0\%$\\
$0.2$ & $3.5$ & $2.950936-0.041980 i$ & $2.950936-0.041980 i$ & $0\%$\\
$0.4$ & $0.1$ & $0.205529-0.164239 i$ & $0.204337-0.166445 i$ & $0.953\%$\\
$0.4$ & $0.5$ & $0.414928-0.037260 i$ & $0.414983-0.038158 i$ & $0.216\%$\\
$0.4$ & $1.$ & $0.830289-0.040368 i$ & $0.830288-0.040335 i$ & $0.00394\%$\\
$0.4$ & $1.5$ & $1.244956-0.040786 i$ & $1.244957-0.040785 i$ & $0.00009\%$\\
$0.4$ & $2.$ & $1.659690-0.040936 i$ & $1.659690-0.040936 i$ & $0\%$\\
$0.4$ & $2.5$ & $2.074460-0.041005 i$ & $2.074460-0.041005 i$ & $0\%$\\
$0.4$ & $3.$ & $2.489252-0.041041 i$ & $2.489252-0.041041 i$ & $0\%$\\
$0.4$ & $3.5$ & $2.904056-0.041063 i$ & $2.904056-0.041063 i$ & $0\%$\\
$0.6$ & $0.1$ & $0.174221-0.128528 i$ & $0.173621-0.129813 i$ & $0.655\%$\\
$0.6$ & $0.5$ & $0.403842-0.035805 i$ & $0.403639-0.037337 i$ & $0.381\%$\\
$0.6$ & $1.$ & $0.807956-0.039107 i$ & $0.807982-0.039085 i$ & $0.00416\%$\\
$0.6$ & $1.5$ & $1.211697-0.039454 i$ & $1.211698-0.039455 i$ & $0.00005\%$\\
$0.6$ & $2.$ & $1.615455-0.039576 i$ & $1.615455-0.039576 i$ & $0\%$\\
$0.6$ & $2.5$ & $2.019233-0.039631 i$ & $2.019233-0.039631 i$ & $0\%$\\
$0.6$ & $3.$ & $2.423024-0.039661 i$ & $2.423024-0.039661 i$ & $0\%$\\
$0.6$ & $3.5$ & $2.826821-0.039679 i$ & $2.826821-0.039679 i$ & $0\%$\\
$0.8$ & $0.1$ & $0.141955-0.102653 i$ & $0.141463-0.103524 i$ & $0.571\%$\\
$0.8$ & $0.5$ & $0.388651-0.034708 i$ & $0.388100-0.035827 i$ & $0.319\%$\\
$0.8$ & $1.$ & $0.777212-0.037481 i$ & $0.777224-0.037492 i$ & $0.00204\%$\\
$0.8$ & $1.5$ & $1.165772-0.037762 i$ & $1.165772-0.037762 i$ & $0.00002\%$\\
$0.8$ & $2.$ & $1.554326-0.037854 i$ & $1.554326-0.037854 i$ & $0\%$\\
$0.8$ & $2.5$ & $1.942886-0.037896 i$ & $1.942886-0.037896 i$ & $0\%$\\
$0.8$ & $3.$ & $2.331449-0.037919 i$ & $2.331449-0.037919 i$ & $0\%$\\
$0.8$ & $3.5$ & $2.720013-0.037932 i$ & $2.720013-0.037932 i$ & $0\%$\\
\hline
\end{tabular}
\caption{Quasinormal modes of the $\ell=0$ test scalar field  for the Yang-Mills black hole ($r_h=1$) $\Lambda=0.001, ~ \xi=0.1$ calculated using the WKB formula at different orders and Pade approximants.}\label{table3}
\end{table}
\begin{table}
\begin{tabular}{c c c c c}
\hline
$Q$ & $\mu$ & WKB-6 ($m=3$) & WKB-9 ($m=4$) & difference  \\
\hline
$0$ & $0.1$ & $0.582204-0.193208 i$ & $0.582212-0.193205 i$ & $0.00135\%$\\
$0$ & $0.5$ & $0.636317-0.160106 i$ & $0.636378-0.160063 i$ & $0.0114\%$\\
$0$ & $1.$ & $0.902468-0.033440 i$ & $0.898956-0.039902 i$ & $0.814\%$\\
$0$ & $1.5$ & $1.294299-0.039276 i$ & $1.294259-0.039277 i$ & $0.00310\%$\\
$0$ & $2.$ & $1.711518-0.040932 i$ & $1.711521-0.040934 i$ & $0.00020\%$\\
$0$ & $2.5$ & $2.131730-0.041524 i$ & $2.131730-0.041524 i$ & $5.8\times 10^{-6}\%$\\
$0$ & $3.$ & $2.553233-0.041808 i$ & $2.553233-0.041808 i$ & $0\%$\\
$0$ & $3.5$ & $2.975425-0.041968 i$ & $2.975425-0.041968 i$ & $0\%$\\
$0.2$ & $0.1$ & $0.570736-0.184292 i$ & $0.570710-0.184260 i$ & $0.00695\%$\\
$0.2$ & $0.5$ & $0.625302-0.151770 i$ & $0.625234-0.151784 i$ & $0.0109\%$\\
$0.2$ & $1.$ & $0.896122-0.028944 i$ & $0.891681-0.038309 i$ & $1.16\%$\\
$0.2$ & $1.5$ & $1.286785-0.039154 i$ & $1.286748-0.039123 i$ & $0.00375\%$\\
$0.2$ & $2.$ & $1.702021-0.040677 i$ & $1.702023-0.040679 i$ & $0.00015\%$\\
$0.2$ & $2.5$ & $2.120116-0.041235 i$ & $2.120116-0.041235 i$ & $3.9\times 10^{-6}\%$\\
$0.2$ & $3.$ & $2.539450-0.041504 i$ & $2.539450-0.041504 i$ & $0\%$\\
$0.2$ & $3.5$ & $2.959448-0.041657 i$ & $2.959448-0.041657 i$ & $0\%$\\
$0.4$ & $0.1$ & $0.535275-0.159460 i$ & $0.535114-0.159465 i$ & $0.0288\%$\\
$0.4$ & $0.5$ & $0.591530-0.128372 i$ & $0.591321-0.128643 i$ & $0.0566\%$\\
$0.4$ & $1.$ & $0.867632-0.031753 i$ & $0.865611-0.032233 i$ & $0.239\%$\\
$0.4$ & $1.5$ & $1.264423-0.038898 i$ & $1.264615-0.038696 i$ & $0.0221\%$\\
$0.4$ & $2.$ & $1.673872-0.039947 i$ & $1.673872-0.039948 i$ & $0.00006\%$\\
$0.4$ & $2.5$ & $2.085612-0.040416 i$ & $2.085612-0.040416 i$ & $0\%$\\
$0.4$ & $3.$ & $2.498461-0.040647 i$ & $2.498461-0.040647 i$ & $0\%$\\
$0.4$ & $3.5$ & $2.911906-0.040780 i$ & $2.911906-0.040780 i$ & $0\%$\\
$0.6$ & $0.1$ & $0.473764-0.126707 i$ & $0.473762-0.126773 i$ & $0.0135\%$\\
$0.6$ & $0.5$ & $0.535136-0.096522 i$ & $0.534823-0.097033 i$ & $0.110\%$\\
$0.6$ & $1.$ & $0.836279-0.033833 i$ & $0.835375-0.038981 i$ & $0.624\%$\\
$0.6$ & $1.5$ & $1.228775-0.037893 i$ & $1.228777-0.037908 i$ & $0.00121\%$\\
$0.6$ & $2.$ & $1.627881-0.038808 i$ & $1.627881-0.038809 i$ & $0.00001\%$\\
$0.6$ & $2.5$ & $2.029042-0.039165 i$ & $2.029042-0.039165 i$ & $0\%$\\
$0.6$ & $3.$ & $2.431139-0.039346 i$ & $2.431139-0.039346 i$ & $0\%$\\
$0.6$ & $3.5$ & $2.833748-0.039451 i$ & $2.833748-0.039451 i$ & $0\%$\\
$0.8$ & $0.1$ & $0.394192-0.099418 i$ & $0.394194-0.099451 i$ & $0.00824\%$\\
$0.8$ & $0.5$ & $0.458860-0.056199 i$ & $0.470506-0.062135 i$ & $2.83\%$\\
$0.8$ & $1.$ & $0.800388-0.034413 i$ & $0.800517-0.034207 i$ & $0.0304\%$\\
$0.8$ & $1.5$ & $1.180130-0.036701 i$ & $1.180128-0.036703 i$ & $0.00021\%$\\
$0.8$ & $2.$ & $1.564856-0.037307 i$ & $1.564856-0.037307 i$ & $0\%$\\
$0.8$ & $2.5$ & $1.951227-0.037558 i$ & $1.951227-0.037558 i$ & $0\%$\\
$0.8$ & $3.$ & $2.338363-0.037688 i$ & $2.338363-0.037688 i$ & $0\%$\\
$0.8$ & $3.5$ & $2.725921-0.037765 i$ & $2.725921-0.037765 i$ & $0\%$\\
\hline
\end{tabular}
\caption{Quasinormal modes of the $\ell=1$ test scalar field  for the Yang-Mills black hole ($r_h=1$) $\Lambda=0.001, ~ \xi=0.1$ calculated using the WKB formula at different orders and Pade approximants.}\label{table4}
\end{table}
\end{tiny}

\section{Quasinormal modes}\label{sec:results}

We distinguish two qualitatively different cases: the asymptotically de Sitter case and the asymptotically flat case. For the latter, we employ the Leaver method, while in the asymptotically de Sitter case with a nonzero field mass, the WKB method is remarkably accurate, as demonstrated in numerous works~\cite{Fontana:2020syy,Dubinsky:2024hmn,Bolokhov:2024ixe,Lutfuoglu:2025hjy}.

The Leaver method requires fixing the radius of the event horizon. We adopt units in which \( r_{h} = 1 \), allowing the black hole mass to be expressed in terms of the remaining parameters \( Q \), \( \Lambda \), and \( \xi \) as:
\begin{equation}
M = \frac{1}{6} \left(6 \xi Q^2 + 3 Q^2 - \Lambda + 3\right).
\end{equation}
For the metric to represent a black hole, the metric function \( f(r) \) must remain positive between the event and cosmological horizons. The parameter region corresponding to valid black hole solutions is shown in Fig.~\ref{fig:BHrange}.

In Fig.~\ref{fig:QNM1}, we observe that the damping rate of the fundamental mode for \( \ell = 0, 1, 2 \) vanishes at a critical value of the scalar field mass \( \mu \), specific to each \( \ell \). When \( \mu \) exceeds this threshold, the fundamental mode disappears from the spectrum and the first overtone becomes dominant. These arbitrarily long-lived modes are known as \textit{quasi-resonances}, first observed in the Schwarzschild background in~\cite{Ohashi:2004wr}.

As illustrated in Figs.~\ref{fig:QNM1} and \ref{fig:QNM2}, such quasi-resonances appear for various values of charge \( Q \) and coupling \( \xi \). The threshold for the emergence of quasi-resonances depends on all parameters of the configuration: \( Q \), \( \xi \), and \( \ell \). We also find that the real part of \( \omega \) increases monotonically with \( \xi \). Interestingly, as the quasi-resonance regime is approached, the quasinormal frequencies for different values of \( \xi \) tend to merge, indicating a form of universal behavior independent of the coupling.

Since the Leaver method is based on a convergent series and is highly accurate, we do not cross-check it with other methods. However, it does require a good initial guess for the frequency, for which we used WKB estimates. It is important to note that in asymptotically flat spacetimes, massive fields are associated with effective potentials that have three turning points~\cite{Galtsov:1991nwq}, whereas the WKB method assumes a potential with only two turning points. Consequently, the relative error of the WKB approximation increases with \( \mu \), and beyond a certain threshold, the method becomes inapplicable due to the absence of a maximum in the effective potential.

The case of a positive cosmological constant is fundamentally different. Here, the effective potential typically exhibits a single peak, monotonically decreasing toward both the event and cosmological horizons, making it well-suited for the WKB approach. 
Indeed, from tables \ref{table1}-\ref{table4} we see that the difference between two quite different WKB orders, sixth and ninth, is negligibly small, being much smaller than the observed effect, that is, the shift of the frequency from its Schwarzschild or Schwarzschild-de Sitter limit. 

While it is generally expected that the quasinormal spectrum of a compact object should not strongly depend on the cosmological background~\cite{Zhidenko:2003wq,Konoplya:2021ube,Konoplya:2005sy}, this is not always the case. In particular, asymptotically de Sitter black holes are an exception~\cite{Konoplya:2004uk,Konoplya:2007zx,Konoplya:2017ymp} where an additional branch of quasinormal modes appears and dominates the late-time signal or even a dynamical instability takes place \cite{Konoplya:2013sba,Cuyubamba:2016cug}.

This leads to one of the most striking features of asymptotically de Sitter spacetime: quasinormal modes govern signal decay at all times~\cite{Dyatlov:2010hq,Dyatlov:2011jd,Dubinsky:2024hmn,Dubinsky:2024gwo,Konoplya:2025mvj}. In contrast, in asymptotically flat spacetimes, the ringdown phase transitions to a late-time power-law tail, and the quasinormal modes do not form a complete basis~\cite{review1,review3}.

Another notable feature is the emergence of a second branch of modes, which are deformations of the empty de Sitter modes due to the presence of the black hole. When the black hole radius is much smaller than the cosmological horizon, the quasinormal modes of this de Sitter branch can be approximated by~\cite{Konoplya:2022kld}:
\begin{equation}
\omega_n = \omega^{(\text{dS})}_n \left( 1 - \frac{M}{r_c} + \mathcal{O}\left( \frac{M}{r_c} \right)^2\right),
\end{equation}
where the (purely imaginary) frequencies \( \omega^{(\text{dS})}_n \) for massless and massive fields in pure de Sitter space are given in~\cite{Lopez-Ortega:2006aal,Lopez-Ortega:2007vlo,Lopez-Ortega:2012xvr}.

This de Sitter branch of modes does not follow the correspondence between null geodesics and eikonal quasinormal modes~\cite{Cardoso:2008bp}, as shown in~\cite{Konoplya:2022gjp}. On the other hand, the Schwarzschild-like branch typically does satisfy this correspondence, as demonstrated in numerous studies~\cite{Konoplya:2020bxa,Bolokhov:2023dxq,Malik:2024qsz,Zinhailo:2018ska,Dubinsky:2024aeu}.

From Tables~\ref{table1} and \ref{table2}, we see that as the field mass increases, the damping rate approaches a constant. The damping in the asymptotically de Sitter case is slower than in the asymptotically flat case, but unlike the latter, it never vanishes regardless of the field mass. In the large-\( \mu \) limit, the quasinormal frequencies become independent of the multipole number \( \ell \), as the dominant mass term suppresses the angular contribution. The absence of quasi-resonances in asymptotically de Sitter black holes has been analytically proven in~\cite{Konoplya:2004wg}.

In the regime of large mass $\mu$ and small $Q$ and $\xi$ we can expand the position of the maximum in terms of powers of $1/\mu$ and further use it in the WKB formula, in a similar manner with \cite{Malik:2025ava}. Then we find that the position of the maximum has the form
\begin{widetext}
\begin{equation}
r_{max} = -\frac{3 M}{x^2-1}-\frac{Q^2}{3
   M}+\xi  \left(\frac{4
   \left(x^2-1\right)^3 Q^2}{27 M^3}\right)+\mathcal{O}\left(\frac{1}{\mu^2}, Q^3, \xi^2\right).
\end{equation}
The quasinormal modes have the form:
\begin{multline}
\omega_n = Q^2 \left(\frac{i (2 n+1) \left(2 x^2-1\right)
   \left(1-x^2\right)^{5/2}}{108 M^3
   x}+\frac{\left(x^2-1\right)^2 \mu }{18 M^2 x}+\xi 
   \left(-\frac{\left(x^2-1\right)^5 \mu }{81 M^4
   x}\right.\right. \\
\left.\left.+\frac{i (2 n+1) \left(1-x^2\right)^{11/2} \left(8
   x^2-1\right)}{486 M^5 x}\right)\right)
   -\frac{i (2 n+1) x
   \left(1-x^2\right)^{3/2}}{6 M}+x \mu 
   +\mathcal{O}\left(\frac{1}{\mu^2}, Q^3, \xi^2\right).
\end{multline}
\end{widetext}
Here a new quantity $x$ is defined as follows:
\begin{equation}
x = \sqrt{1- \sqrt[3]{9M^2\Lambda }}.
\end{equation}
From the above analytic expression we see that in the regime of large $\mu$ quasinormal modes indeed do not depend on the multipole number $\ell$, which is in concordance with the data presented in tables \ref{table1}-\ref{table4}. The above analytical expression can, in principle, be extended to higher orders; however, even the next term in the \( 1/\mu \) expansion becomes quite cumbersome.

Finally, we examined the stability of scalar field perturbations with \( \ell = 0 \), both for massless fields and for fields with a small nonzero mass, in cases where the effective potential develops a negative gap. In our computations, we did not observe any growing modes, indicating stability in all considered cases. However, for a massless scalar field in a near-extremal black hole background—where the negative gap becomes relatively deep—we observed that the late-time dominant de Sitter mode is purely imaginary and exhibits a very small damping rate (see Fig.~\ref{fig:TD1}). It is worth noting that asymptotically flat black holes do not exhibit a negative gap in the effective potential and are thus manifestly stable.

\section{CONCLUSIONS}\label{sec:conclusion}

In this work, we have investigated the quasinormal modes of a massive scalar field in the background of non-minimal Einstein–Yang–Mills black holes with a non-zero cosmological constant. Using the Leaver method for asymptotically flat spacetimes and the WKB method for asymptotically de Sitter cases, we have demonstrated several distinct features in the quasinormal spectrum that emerge due to the interplay between the field mass, the non-minimal coupling, and the cosmological constant.

For asymptotically flat spacetimes, we confirmed the existence of quasi-resonances—arbitrarily long-lived modes that arise when the scalar field mass exceeds a certain critical value. These modes are particularly sensitive to the multipole number and field mass. In contrast, for asymptotically de Sitter spacetimes, we found that the damping rate remains finite regardless of the field mass, confirming the absence of quasi-resonances in this regime. The quasinormal frequencies in this case become independent of the angular momentum at large masses, due to the suppression of the centrifugal term by the dominant mass term.

Interestingly, as shown in Fig.~\ref{fig:QNM2}, the threshold for the onset of quasi-resonances is independent of the non-minimal coupling \( \xi \); the quasinormal frequencies for different values of \( \xi \) merge as the damping rate approaches zero, indicating a universal behavior in the quasi-resonant regime.

In the asymptotically de Sitter case, massive fields have been shown to obey, with high accuracy, the correspondence between grey-body factors and quasinormal modes~\cite{Konoplya:2024lir,Malik:2024cgb}. Therefore, the quasinormal modes obtained in this paper can be directly used to compute the corresponding grey-body factors in similar way with \cite{Dubinsky:2024vbn}. Furthermore, the present analysis could be extended to fields with higher spin.

\vspace{4mm}
\begin{acknowledgments}
The author thanks R. A. Konoplya for useful discussions. The author acknowledges the University of Seville for their support through the Plan-US of aid to Ukraine.
\end{acknowledgments}

\bibliography{bibliography}
\end{document}